\documentclass{aastex}
\usepackage{spr-astr-addons}

\RequirePackage{color}

\begin{document}

\title{First Stars. II. Evolution with mass loss}

\slugcomment{}
\shorttitle{}
\shortauthors{}

\author{D. Bahena\altaffilmark{1}} \and
\author{P. Hadrava\altaffilmark{1}}

\altaffiltext{1}{Astronomical Institute of the Academy of Sciences,\\
              Bo\v{c}n\'{\i} II 1401, 14131 Praha 4, Czech Republic.\\
              e-mail: bahen@hotmail.com, had@sunstel.asu.cas.cz}

\begin{abstract}
The first stars are assumed to be predominantly massive. Although,
due to the low initial abundances of heavy elements the line-driven
stellar winds are supposed to be inefficient in the first stars,
these stars may loose a significant amount of their initial mass by other
mechanisms.
In this work, we study the evolution with a prescribed mass loss
rate of very massive, galactic and pregalactic, Population III
stars, with initial metallicities $Z=10^{-6}$ and $Z=10^{-9}$,
respectively, and initial masses $100$, $120$, $150$, $200$, and
$250\,M_{\odot}$ during the hydrogen and helium burning phases.
The evolution of these stars depends on their initial mass,
metallicity and the mass loss rate. Low metallicity stars are
hotter, compact and luminous, and they are shifted to the blue upper
part in the Hertzprung-Russell diagram. With mass loss these stars
provide an efficient mixing of nucleosynthetic products, and
depending on the He-core mass their final fate could be either
pair-instability supernovae or energetic hypernovae. These stars
contributed to the reionization of the universe and its enrichment
with heavy elements, which influences the subsequent star formation
properties.
\end{abstract}

\keywords{first stars, stars: models, evolution, mass loss}


\section{Introduction}
\label{sec:introduction}


It is widely accepted that the so called Population III stars, i.e.
the first stars formed from the primordial gas with an extremely low
metallicity, have played an important role in the evolution of the
universe. The first stars are assumed to be predominantly very
massive and they were responsible for reionization of the universe
and its enrichment by heavy elements.

The formation of the first stars and hence also their Initial Mass
Function (IMF) is significantly influenced by the low metallicity of
the protostellar material. Due to its low opacity, its gravitational
attraction can not be effectively opposed by radiative pressure and
the continuing accretion gives rise to more massive stars.

Numerical hydrodynamic simulations of star formation generally
confirm the tendency to higher masses of the first stars, although
quantitatively the results differ considerably.

\cite{Bromm2004} found the masses of the first stars between
$10^{2}$ and $10^{3}M_{\odot}$, \cite{Omukai2003} followed the
evolution of accreting protostars and found $\sim 600\,M_{\odot}$ as
the upper limit for massive stars. According to \cite{Abel2002}, the
final masses are uncertain because a single molecular protostar seed
of $\sim 1 M_{\odot}$ at the centre of a $\sim 100\,M_{\odot}$ core
of a protogalaxy is formed. The cores may fragment into clusters of
massive (MS) or very massive stars (VMS). \cite{Nakamura1999} found
$3\,M_{\odot}$ as the mass of the first stars which may grow up to
$\sim 16\,M_{\odot}$ by the accretion. The IMF of Population III
could thus be bimodal with peaks at $\sim 1-2\,M_{\odot}$ and $\sim
10^{2}\,M_{\odot}$ \citep{Nakamura2001}. \cite{Tumlinson2004}
defined \textit{strong VMS hypothesis} (``The first generation were
exclusively VMS") and the \textit{weak VMS hypothesis} (``The first
generation included VMS in addition to MS with $M\lesssim
50\,M_{\odot}$").

The structure, evolution and properties of the first stars have been
modelled in several studies including the Paper\,I of this series
\citep{Bahena2010}. The results showed that the Population III stars
are smaller but more luminous and hotter than the normal
metal-richer Population I an II stars of equal masses.

The massive luminous normal stars have a high mass loss mainly due
to the stellar wind driven by radiative pressure in lines of metals
\citep[the so called CAK-wind]{Castor1975}. This process cannot take
place in the metal-free stars \citep{Bromm2003}. This is why
practically all models of first-stars evolution have neglected
completely the mass loss.

Several recent studies, however, revealed that even in the absence
of CAK-wind there may occur a considerable mass loss due to
different alternative processes.

A reliable quantitative theory of the mass loss rate in the
Population III stars is still missing, but it is of great interest
to investigate the possible consequences of the mass loss on the
properties of these stars. For this purpose we extend hereby our
calculations of the conservative case of the first-stars evolution
in Paper\,I assuming a mass loss rate given by an ad-hoc
parametrization.

This work is organized as follows: In Sect.\,\ref{massloss} the mass
loss mechanisms are reviewed. In Sect.\,\ref{results} we describe
the initial conditions and results of the computed models, the main
physical variables and their properties. Then, in
Sect.\,\ref{discussion} we discuss our results in comparison with
the conservative case. Finally, in Sect.\,\ref{conclusions} we
summarize the conclusions.


\section{Mass Loss Mechanisms \label{massloss}}


The mass loss in stars of Population I and II has been studied both
theoretically and observationally. However, an extrapolation of the
results to stars of Population III is questionable due to very
different physical properties of both the interiors as well as the
atmospheres of the stars.

Regarding the short lifetimes of the massive first stars, their mass
loss rates have thus to be estimated from the theoretical
considerations and only an indirect evidence can be gained from
observations of later generations of stars. E.g., the abundances of
elements and isotopes observed in metal-poor halo stars
\citep{Beers2005} suggest that these stars were formed from an
interstellar matter enriched by mass lost during the H- and He-
burning in the first stars rather than by debris of pair-instability
supernovae (SNe) expected to be the final stage of the first VMS
evolution \citep{Hirschi2007,Cescutti2010}.

\subsection{Stellar winds}

The mass loss via stellar winds influences significantly the
evolution of massive Population I and II stars and contributes a
great deal to the enrichment of interstellar matter by heavy
elements.

For cool winds of Population I red giants on the RGB
\cite{Reimers1975, Reimers1977} found an empirical relation giving
the mass loss rate as

\begin{equation}\label{Reimers}
 \dot{M}\sim \frac{L}{gR}\sim \frac{LR}{M}\; ,
\end{equation}

\noindent where $L$, $R$, $M$ and $g$ are the luminosity, radius,
mass and surface gravity of the star. \cite{Schroder2005} obtained a
semi-empirical generalization of the Reimers' relation.

 The stellar winds are significantly enhanced at the early-type stars
by radiative pressure either in continuum or especially in lines.
The dependence of the mass loss rate due to CAK-wind on the
metallicity has been scaled by the relation

\begin{equation}\label{Zz}
  \dot{M}\sim Z^{\zeta}\; ,
\end{equation}

\noindent where the exponent varies between $0.5-0.8$ for $0.01\leq
Z/Z_{\odot}\leq 1.0$ \citep{Leitherer1992, Vink1999, Vink2000,
Vink2001}. However, \cite{Kudritzki2002} has found that this
relation breaks down below a certain threshold of the metallicity.

 \cite{Krticka2006, Krticka2009, Krticka2010b} have studied
line-driven winds from the first stars. They showed that weak
stellar winds may be due to absorption in lines of CNO elements
carried by mixing from the interiors to the atmospheres of the
stars.

\subsection{Rotation}

The Population III stars are supposed to be faster rotators than the
stars with higher metallicity \citep{Maeder1999,Meynet2006b}. The
numerical hydrodynamic models of star-formation by \cite{Stacy2011}
also indicate high angular momenta of the first stars, what
influences their structure, evolution and final fate. The effects of
rotation on the structure and evolution of massive stars were
studied by the Geneva Group \citep{Meynet1997, Maeder1998,
MaederMeynet2001, Meynet2000, Meynet2002}. The rotation may increase
the mass loss by several processes:

\begin{enumerate}
\item The rotational mixing enhances the atmospheres with heavier
elements which facilitate the radiatively driven winds
\citep{Lamers1995}.

\item The gravitational darkening enhances the radiative flux and the
mass loss on poles of the rotationally oblate stars while reducing
the angular-momentum loss \citep{MaederDesjacques2001,
Dwarkadas2002}.

\item The centrifugal force reduces the gravity on the equator. When
approaching the critical rotation, the decretion disks are formed
through which the angular momentum and significant amount of mass is
lost \citep{Krticka2010a,Krticka2011}.
\end{enumerate}

\subsection{Pulsation}

The pulsational instability of very massive stars might contribute
somewhat to the total mass loss rate although it is not clear what
the contribution will be. Cool supergiants do indeed have winds
which are initiated by pulsations, but then further support, such as
radiative force on the dust grains, is needed to sustain a
significant outflow.

 The stability of very massive stars was analyzed and compared with
metal enriched stars by \cite{Baraffe2001}. Mass loss was not taken
into account because they assumed it to be negligible for
zero-metallicity stars. A linear non-adiabatic analysis was carried
out. In their analysis, they considered that the stars stabilize as
soon as they evolve from the ZAMS, then one can expect decreasing
mass loss rates along the main sequence. This is, pulsations do not
lead to significant mass loss on the main sequence. For their
calculations they used the so called {\it flux-freezing convective
theory} that assumes that perturbation in the luminosity is only
given by the radiative luminosity. However, the stability properties
of the models could change if convection is properly taken into
account \citep{Klapp2005}.

 Thus, the issue of whether very massive stars could be vibrationally
unstable to pulsation and induce pulsationally driven mass loss is
still open. The contribution of pulsation to mass loss remains
uncertain.

\subsection{Other mechanisms}

\cite{Smith2006} suggested that the mass loss during the evolution
of Population III stars may be dominated by optically thick,
continuum-driven outbursts or hydrodynamical explosions, similarly
as in the very massive Luminous Blue Variable (LBV) stars and
$\eta$~Car in particular. Unlike the steady line-driven winds, this
mechanism, investigated also, e.g., by \cite{vanMarle2008a,
vanMarle2008b}, is insensitive to metallicity but requires a high,
nearly Eddington luminosity, close to which the mass loss rates of
the wind highly increase \citep{Kudritzki2002, Grafener2008,
Vink2011}. This condition is well satisfied by the first stars owing
to their high masses and luminosities.

 The role of magnetic fields in activity of late-type stars and
their winds is generally known. The growing observational evidence
of magnetic fields in early-type stars (e.g. Be-stars) suggests that
they could also be present in the first stars and consequently
influence their winds or mass loss via their decretion disks
\citep[cf., e.g.,][ and references therein]{Puls2008,udDoula2008}.
However, it is not clear whether there was a magnetic field in the
interstellar medium in the early universe.

 Finally, because the first stars were also created in binaries
and multiple systems, they could be subjected to mass-exchange, mass
loss from their systems and enhancement of their winds by tidal
forces from companion stars as it is theoretically well understood
and observationally confirmed for the present generations of stars.

 Motivated by these studies we explore a scenario for the evolution
of very massive and low-metallicity Population III mass losing
stars.


\section{Evolutionary Models at Low Metallicity}
\label{results}


\subsection{Input physics and initial conditions}

In agreement with \cite{Castellani2000} we take as the Population
III stars all the stellar objects with metallicity below a not yet
well defined upper limit between $Z=10^{-10}$ and $10^{-6}$.
Depending on their metallicity we distinguish between galactic and
pregalactic stars, the former with metallicity $Z=10^{-6}$, and the
later with $Z=10^{-9}$, both considered as lower metallicity stars,
but not zero-metallicity stars.

 The concept of critical metallicity has been used to characterize
the transition between Population III and Population II star
formation modes \citep{Omukai2001, Bromm2001, Schneider2002,
Schneider2003, Mackey2003, Bromm2004}.

 Our evolutionary models for both galactic and pregalactic Population
III stars have been calculated by using a stellar code described in
Paper I by \cite{Bahena2010}. These models assume the standard
quasistatic non-rotating configuration of the stars. A
semiconvection treatment for convective transport of energy has been
used. In addition to the assumptions on the underlying physics used
in Paper I, we include now also a mass loss parametrization.

 In view of the uncertainty in the dominant mass loss process in
the Population~III stars, the corresponding mass loss rates and
their changes with the stellar parameters as summarized in the
previous Section, we have adopted a simple parametrization of the
mass loss rate, based on the first relations proposed by
\cite{Lucy1970} and \cite{Barlow1977} to explain the mass loss for
hot luminous stars,

\begin{equation}\label{MLR}
 \dot{M}=N L/c^2\; ,
\end{equation}

\noindent{}where $L$ is the luminosity of the star, $c$ is the speed
of light, and $N$ is a mass loss parameter. This parameter is taken
as an average measure of the mass loss rate and it is kept constant
in the course of the whole evolution. It is based on the simplifying
assumption that the mass-energy output of the star in the form of
non-zero rest-mass particles is proportional to its radiative
output.

 Different models of stellar winds give the mass loss rates in the
form of power-law in luminosity with various exponents and they
usually also depend on other stellar parameters. Let us note that
unlike the Reimers' formula (\ref{Reimers}) corresponding to the
stellar winds of late-type stars, which for chosen $M$ gives the
$\dot{M}$ proportional to $R$ increasing for orders of magnitude for
stars evolving from the main sequence to the red-giant branch, our
approximation (\ref{MLR}) intended to an unspecified mass loss
mechanism varies linearly with $L$ only, i.e. much less in the
course of evolution of the star.

It can be expected that the mass loss rate is more complicated in
reality, however, to get an insight into its possible effects on the
stellar evolution before a proper model of the mass loss mechanism
will be available we are limited to the use of such an {\it ad hoc}
assumption. In our parametrization, the mass loss parameter $N=100$
gives for the luminosity $L\sim 10^{6}L_{\odot}$ typical for stars
with mass of the order $100\,M_{\odot}$ a mass loss rate of
$\dot{M}\sim 10^{-5} M_{\odot}$yr$^{-1}$ which is able to reduce
significantly the initial mass of the star during its characteristic
life-time $\tau\sim 10^{6}$yr. Calculations with different $N$
values should be indicative for the way in which the evolution of
massive first stars is affected by the mass loss.

 To facilitate a comparison with models without mass loss the initial
conditions for the models presented here have been chosen the same like
in Paper I.

 The chemical composition corresponding to galactic lower metallicity
stars is ($X,Z=0.765,10^{-6}$), and for pregalactic ones
($X,Z=0.765,10^{-9}$) by using a value $Y=0.235$ as a primordial
helium value. The initial stellar masses have been chosen $100$,
$120$, $150$, $200$, and $250\,M_{\odot}$. For the mass loss
parameter we have used the values $N=50$ and $100$.

\subsection{Properties of the models} \label{properties}

The evolutionary models presented here predict the effective
temperature, luminosity and other physical variables and quantities
for very massive Population III mass losing stars during the
hydrogen and helium burning phases, for the initial masses and mass
loss parameters chosen.

\begin{figure}
\begin{center}
\includegraphics [width=82mm]{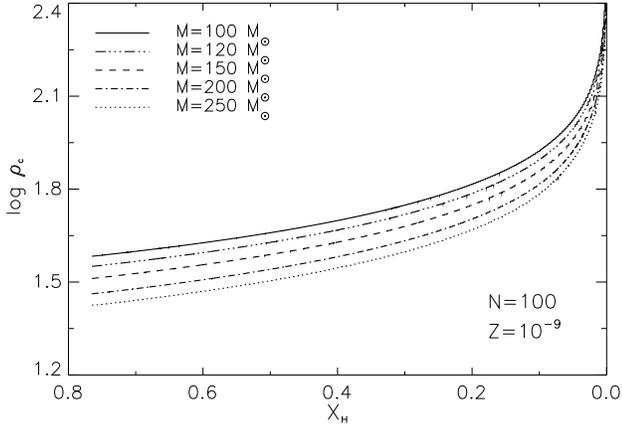}
\caption{Central density for $100\,M_\odot$ (solid line),
$120\,M_\odot$ (dash-dot-dot-dot-dash), $150\,M_\odot$ (dashes),
$200\,M_\odot$ (dash-dot-dash) and $250\,M_\odot$ (dots) pregalactic
Population III stars with metallicity $Z=10^{-9}$ and mass loss
parameter $N=100$ during the hydrogen burning phase.}
\label{bar_fig1}
\end{center}
\end{figure}

\begin{figure}
\begin{center}
\includegraphics [width=82mm]{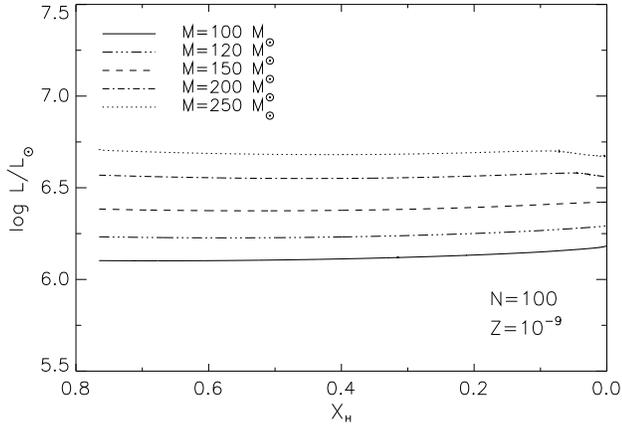}
\caption{Idem dito. Luminosity} \label{bar_fig2}
\end{center}
\end{figure}

\begin{figure}
\begin{center}
\includegraphics [width=82mm]{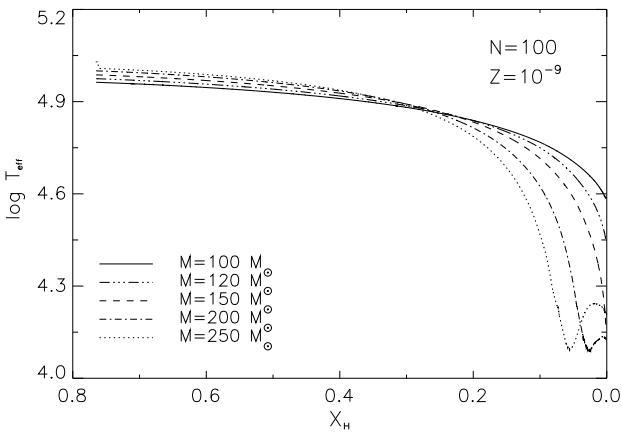}
\caption{Idem dito. Effective temperature} \label{bar_fig3}
\end{center}
\end{figure}

\begin{figure}
\begin{center}
\includegraphics [width=82mm]{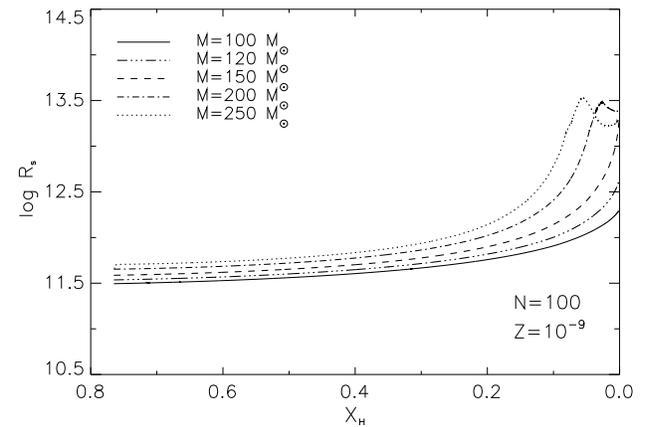}
\caption{Idem dito. Radius} \label{bar_fig4}
\end{center}
\end{figure}

 Figures \ref{bar_fig1} to \ref{bar_fig4} show some physical variables
during the hydrogen and helium burning for pregalactic stars with
masses $100$, $120$, $150$, $200$, and $250\,M_{\odot}$, metallicity
$Z=10^{-9}$ and mass loss parameter $N=100$.

 Population III stars are hotter than their metal enriched
counterparts, and their locus in the HR-diagram is shifted to the
left upper part; pregalactic stars are bluer than the galactic ones.

 For a given stellar mass of main-sequence stars its location in the
HR-diagram depends strongly on metallicity but also on the mass loss
strength. During the hydrogen burning, the higher the mass loss
rate, the greater the reduction in luminosity.

 Our models exhibit a high production of ionizing photons. This
confirms the important cosmological consequences of the first
generation of stars for the reionization of the universe.

Depending on their initial mass and the mass loss rate the final
fate of our models could be as supernovae or hypernovae explosions.

\subsubsection*{\qquad a) Central density and temperature}

Lower metallicity stars are denser than the enriched stars. At the
beginning of the evolution, galactic and pregalactic stars have
different central densities which increase with decreasing $Z$. The
central density increases with decreasing stellar mass and it is
higher for lower metallicity stars.

 As the evolution goes on, the mean molecular weight increases and
the density and temperature increase. During hydrogen burning, both
the central density and temperature increase slowly until hydrogen
is exhausted in the core.

 The most massive stars, and lower metallicity stars, have higher
central temperature. According to \cite{Bromm2001}
the central temperature is related to the mass and radius as

\begin{equation}
    T_{\rm c}\sim\frac{M^{1/2}}{R}\; .
\end{equation}

For stars with $M>100\,M_{\odot}$ the central temperature is
insensitive to both mass and radius and depends upon the composition
only. In the absence of metals, $T_{\rm c}$ is higher and the star
self-generates the small amount of CNO elements needed for energy
generation to support the star \citep{Castellani2000}.

\subsubsection*{\qquad b) Effective temperature and radius}

At the beginning of the main sequence, galactic stars with
metallicity $Z=10^{-6}$, and mass loss parameter $N=50$, have an
effective temperature $T_{\rm eff}=69582$, $73416$, and $75586$ K
for a $100$, $150$ and $200\,M_{\odot}$, respectively. For the same
stellar masses, pregalactic stars with $Z=10^{-9}$ have $T_{\rm
eff}=91865$, $97109$, and $100090$ K, respectively. Due to mass
loss, the effective temperatures decrease during core hydrogen
burning. At the end, these temperatures are lower than for the
conservative case.

 On the main sequence, $100$, $150$, and $200\,M_{\odot}$ galactic
stars with metallicity $Z=10^{-6}$, have initial radii $R=7.67$,
$9.54$ and $11.15\,R_{\odot}$, respectively, and for pregalactic
stars with $Z=10^{-9}$, their radii are $R=4.48$, $5.55$ and
$6.46\,R_{\odot}$, respectively.

 With decreasing effective temperature, the radius increases. For
stars evolving with high mass loss parameters, their radii at the
end of the hydrogen burning increase even more.

 The changes of the radius and effective temperature can be
understood in terms of changes in the convective core size. To
satisfy the boundary condition

\begin{equation}
  L=\pi acR^{2}T_{\rm eff}^{4}
\end{equation}

\noindent the effective temperature decreases with increasing
radius. An expanding convective core implies a decreasing radius and
increasing effective temperature.

\subsubsection*{\qquad b) Convective core size}

During the hydrogen burning the convective core evolves in different
ways depending on whether the stars are mass losing or not. If a
star evolves with mass loss its convective core decreases
continuously during the hydrogen burning phase. In the conservative
case, a helium core is formed only at the end of hydrogen burning
when the star is contracting during its transition to helium
burning.

\begin{figure}
\begin{center}
\includegraphics [width=82mm]{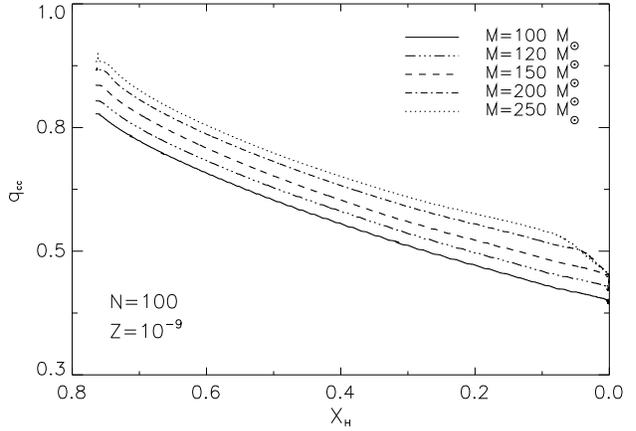}
\caption{Convective core size for $100\,M\odot$ (solid line),
$120\,M_\odot$ (dash-dot-dot-dot-dash), $150\,M_\odot$ (dashes),
$200\,M_\odot$ (dash-dot-dash) and $250\,M_\odot$ (dots) pregalactic
Population III stars with metallicity $Z=10^{-9}$ and mass loss
parameter $N=100$ during the hydrogen burning.}\label{bar_fig5}
\end{center}
\end{figure}

 At the end of hydrogen burning, the convective core size, defined
as the ratio $q_{\rm cc}=M_{\rm cc}/M$ of the convective core mass
$M_{\rm cc}$ to the total stellar mass $M$, is larger for evolution
with mass loss than for the conservative case. This is because for
mass losing stars their initial mass is reduced during the
evolution. Figure\,\ref{bar_fig5} shows the evolution of the
convective core size during the hydrogen burning for pregalactic
Population III stars and mass loss parameter $N=100$.

\subsubsection*{\qquad d) Luminosity and Gamma factors}

Luminosity increases or decreases during evolution with mass loss
depending on the mass loss rate. The rate of decrease of the
luminosity is higher for large mass loss parameters.

\begin{figure}
\begin{center}
\includegraphics [width=82mm]{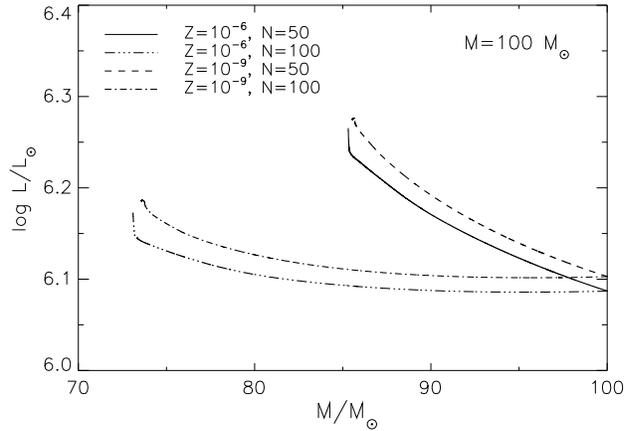}
\caption{Mass-luminosity relation for $100\,M_\odot$ galactic and
pregalactic Population III stars with metallicity $Z=10^{-6}$ and
$Z=10^{-9}$, respectively, and mass loss parameters $N=50$ and
$N=100$, during the hydrogen burning.}\label{bar_fig6}
\end{center}
\end{figure}

 In Figure~\ref{bar_fig6}, the mass-luminosity ($M-L$) relation is shown for
$100\,M_{\odot}$ galactic and pregalactic Population III mass losing
stars during the hydrogen burning. For low values of the mass loss
parameter $N=50$, the luminosity increases during hydrogen burning
but decreases for large values of $N=100$. This is because in VMS
the opacity is mostly due to electron scattering. Then, as the
evolution goes on, the opacity decreases and the mean molecular
weight increases. Both of these effects contribute to an increase of
luminosity which is proportional to the stellar mass. The $M-L$
relation reads

\begin{equation}
  L\sim M^{\eta}\; ,
\end{equation}

\noindent where the value of the exponent $\eta$ depends on the mass
loss rate. The evolution reaches a given mass with higher
luminosity, the lower the value of $N$ is.

\begin{figure}
\begin{center}
\includegraphics [width=82mm]{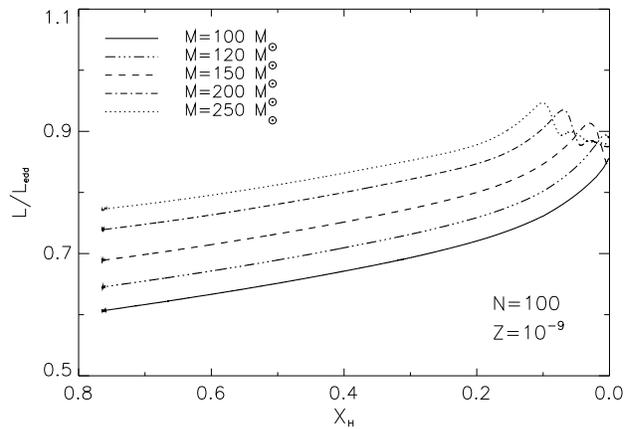}
\caption{The $\Gamma$ factor for pregalactic Population III stars
with metallicity $Z=10^{-9}$ and mass loss parameter $N=100$ during
the hydrogen burning.}\label{bar_fig7}
\end{center}
\end{figure}

 Figure ~\ref{bar_fig7} shows the Gamma factor during the core
hydrogen burning, for pregalactic Population III stars with a mass
loss parameter $N=100$.

 Galactic, and pregalactic Population III stars evolve during
hydrogen burning near the Eddington upper luminosity limit. With and
without mass loss, the ratio between the luminosity of the star and
its Eddington luminosity is similar.

\begin{figure}
\begin{center}
\includegraphics [width=82mm]{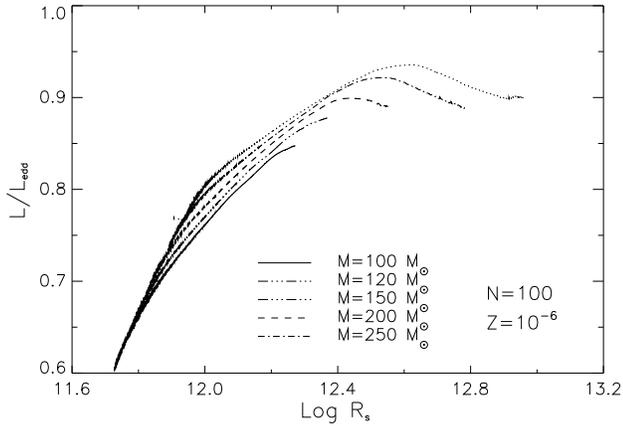}
\caption{The $\Gamma$ factor as function of the radius for galactic
Population III stars with metallicity $Z=10^{-6}$ and mass loss
parameter $N=100$ during the hydrogen burning.}\label{bar_fig8}
\end{center}
\end{figure}

In Figures ~\ref{bar_fig8} and 9, the Gamma factor is plotted as
a function of the radius showing the phase when this factor approaches
to unity, during the hydrogen burning, for mass losing galactic and
pregalactic stars, respectively.

\begin{figure}
\begin{center}
\includegraphics [width=82mm]{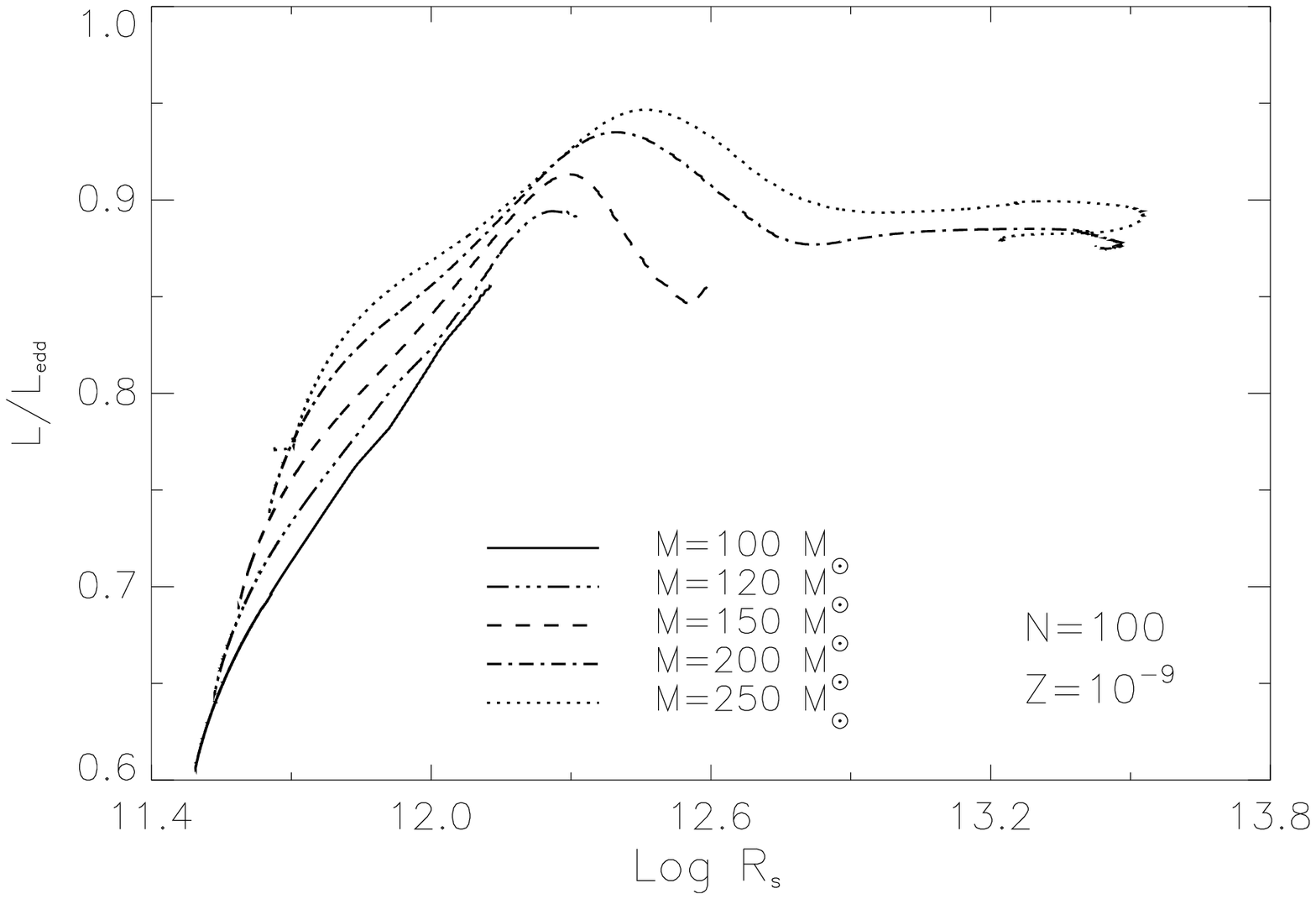}
\caption{Idem dito. Pregalactic Population III stars with
metallicity $Z=10^{-9}$ and mass loss parameter $N=100$ during the
hydrogen burning.}\label{bar_fig9}
\end{center}
\end{figure}

\begin{figure}
\begin{center}
\includegraphics [width=82mm]{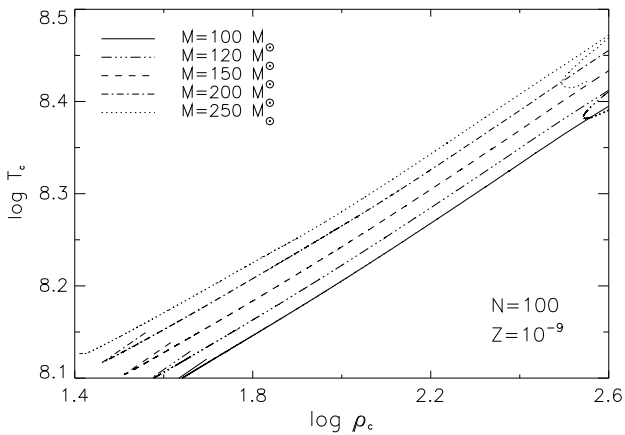}
\caption{${\rm log}\,T-{\rm log}\,\rho$ plane for $100$, $120$,
$150$, $200$ and $250\,M_{\odot}$ pregalactic Population III stars
with metallicity $Z=10^{-9}$, and mass loss parameter $N=100$,
during the hydrogen and helium burning phases.}\label{bar_fig10}
\end{center}
\end{figure}

\subsubsection*{\qquad e) $\rho-T$ plane}

In the $\rho-T$ plane the galactic and pregalactic very massive
Population III mass losing stars occupy the upper loci of a
nondegenerate and nonrelativistic gas. In this zone there is an
upper boundary at which pair production effects could become
important. The most massive stars tend to approach to it, however,
our models are below this boundary. Regarding the gas in
thermodynamic equilibrium, mass losing stars studied here are
dominated by radiation pressure. Fig. \ref{bar_fig10} shows the
evolution of conditions in the centres of stars.

\subsubsection*{\qquad f) HR diagram}

 As in the conservative case, the locus of very massive Population
III stars in the HR diagram is in the left upper part. Pregalactic
stars are hotter than the galactic ones. The most massive stars are
the most luminous. Depending upon the star's initial mass and mass
loss rate, its evolution can be different.

\begin{figure}
\begin{center}
\includegraphics [width=72mm]{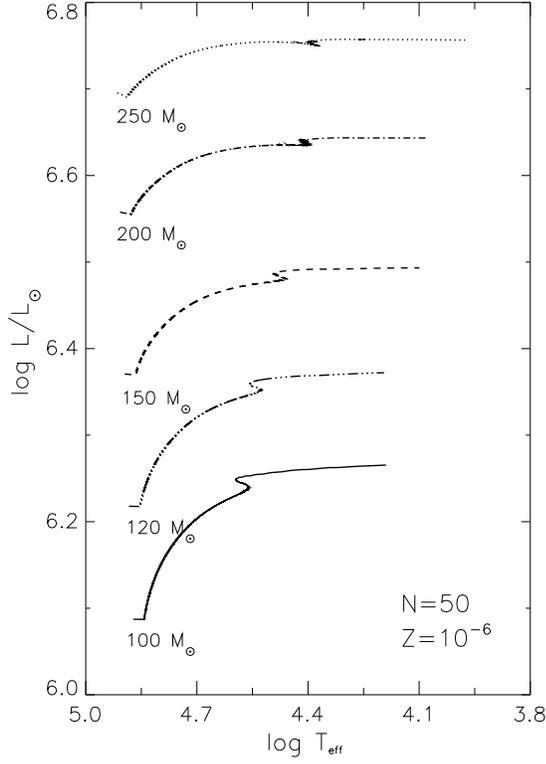}
\caption{Evolutionary tracks in the HR-diagram for $100$, $120$,
$150$, $200$ and $250\,M_{\odot}$ galactic Population III stars with
metallicity $Z=10^{-6}$ and mass loss parameter $N=50$, during the
hydrogen and helium burning phases.} \label{bar_fig11}
\end{center}
\end{figure}

\begin{figure}
\begin{center}
\includegraphics [width=70mm]{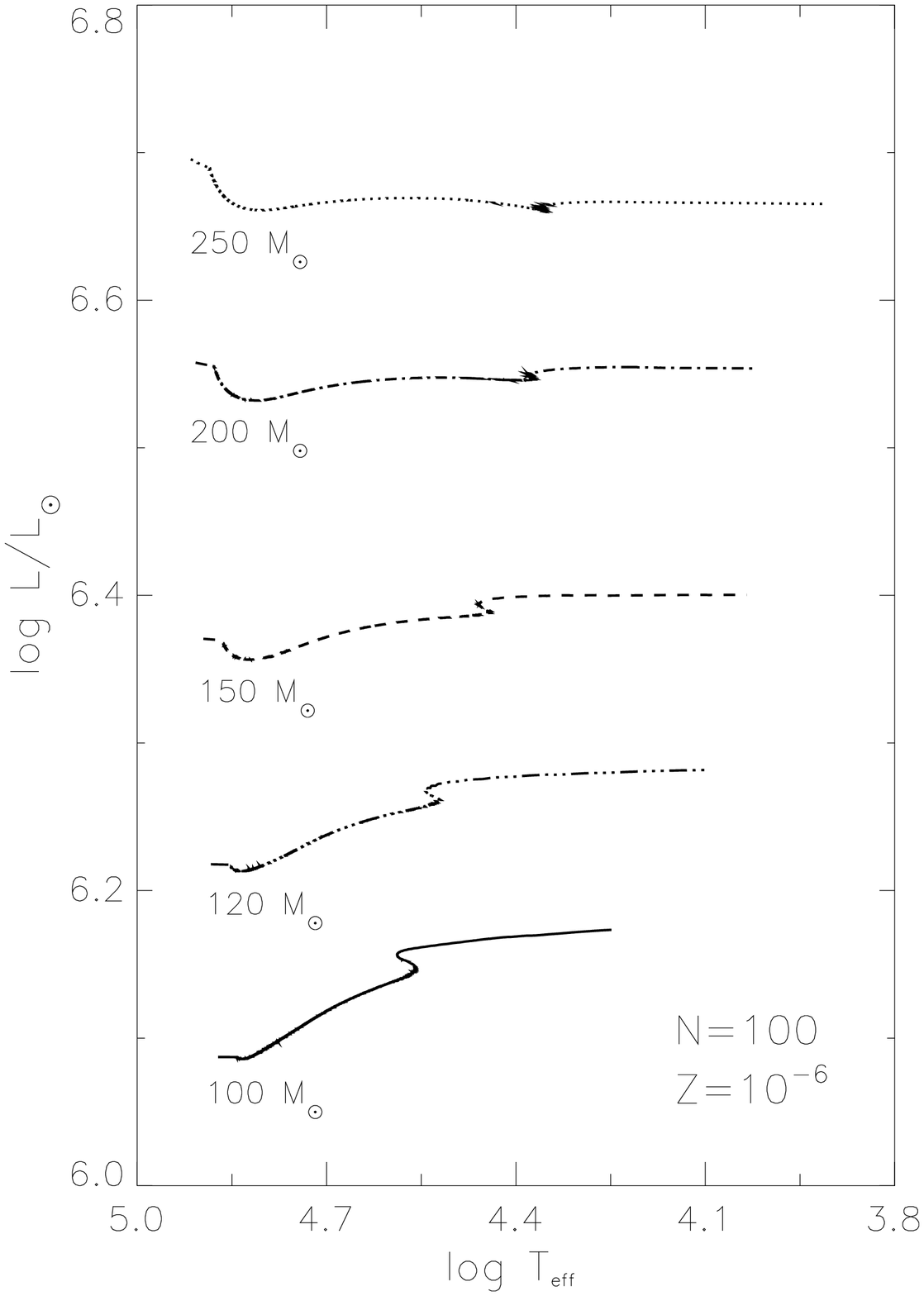}
\caption{Idem dito. Mass loss parameter $N=100$.} \label{bar_fig12}
\end{center}
\end{figure}

\begin{figure}
\begin{center}
\includegraphics [width=72mm]{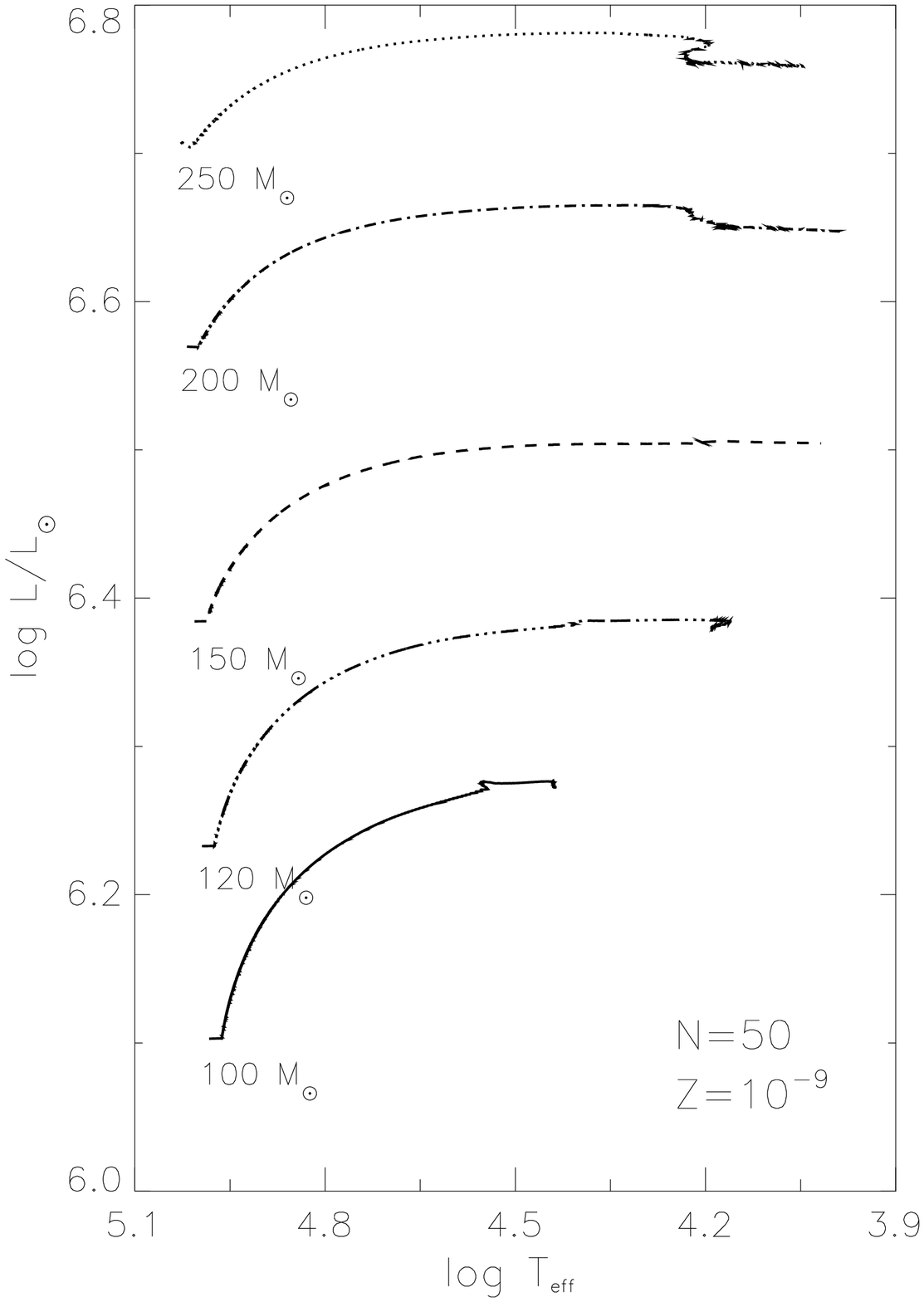}
\caption{Evolutionary tracks in the HR-diagram for $100$, $120$,
$150$, $200$ and $250\,M_{\odot}$ pregalactic Population III stars,
metallicity $Z=10^{-9}$, and mass loss parameter $N=50$, during the
hydrogen and helium burning phases.} \label{bar_fig13}
\end{center}
\end{figure}

\begin{figure}
\begin{center}
\includegraphics [width=70mm]{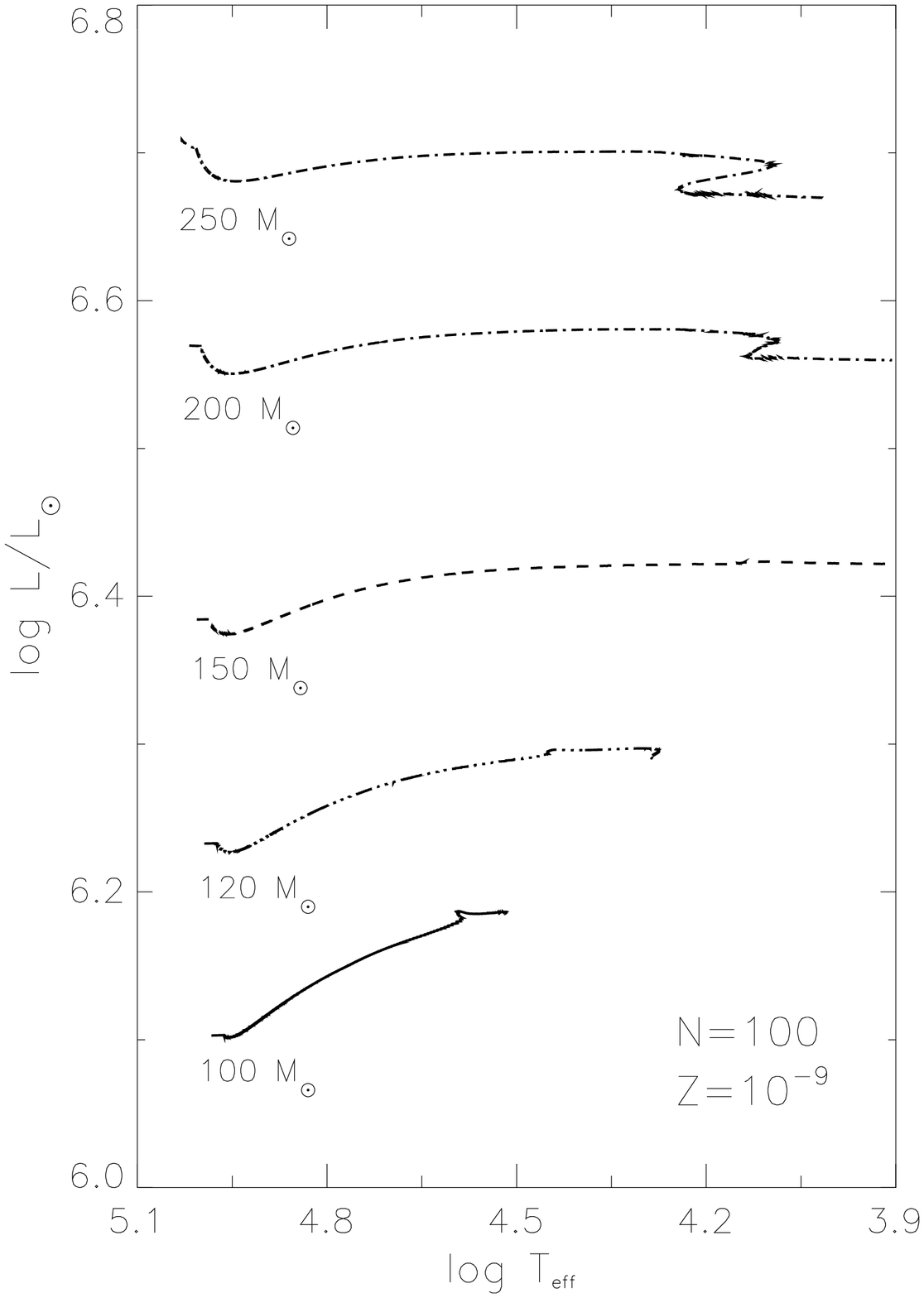}
\caption{Idem dito. Mass loss parameter $N=100$.} \label{bar_fig14}
\end{center}
\end{figure}

 During hydrogen burning, the luminosity of a star without mass loss
increases while a helium core is formed. Then, as the star evolves,
the luminosity remains almost constant while the effective
temperature decreases. However, for different stars their luminosity
and effective temperature decrease with decreasing mass and
increasing metallicity. The luminosity is lower for mass losing
stars than for stars without mass loss.

 Different stars evolve with different lifetimes. Depending on the
mass loss rate, both the initial hydrogen and helium burning phases
take place at the blue side of the HR diagram. When the stars evolve
with assumed mass loss rates they move toward the red, like in the
conservative case. With and without mass loss the stars do not
experience the AGB phase.

Figures ~\ref{bar_fig11} to ~\ref{bar_fig14} show evolutionary
tracks in the HR diagram for $100$, $120$, $150$, $200$ and
$250\,M_{\odot}$, for mass loss parameters $N=50$ and $N=100$,
corresponding to galactic and pregalactic Population III stars with
metallicity $Z=10^{-6}$ and $Z=10^{-9}$, respectively.


\section{Discussion} \label{discussion}


Population III stars are defined as all metal deficient stars which share the
common peculiarity of experiencing self-production of C sometimes
during their H-burning evolution \citep{Castellani2000}. That is,
these stars may build-up a small fraction of C nuclei ($Z_{\rm
C}\sim 10^{-10}$) via the triple-$\alpha$ reactions during the
pre-main sequence \citep{Cassisi1993}.

 Then, these stars produce their nuclear energy during the H-burning phase
from a combination of the proton-proton (pp) chain and the CNO-cycle
with the small amount of C previously built. To provide the
necessary luminosity, the star has to reach very high central
temperatures $T_{\rm c}\sim 10^{8}$ K for the simultaneous
occurrence of the helium burning via the $3\alpha$ process. This
burning produces certain amount of heavy elements to follow
efficiently the CNO-cycle.

\begin{figure}
\begin{center}
\includegraphics [width=82mm]{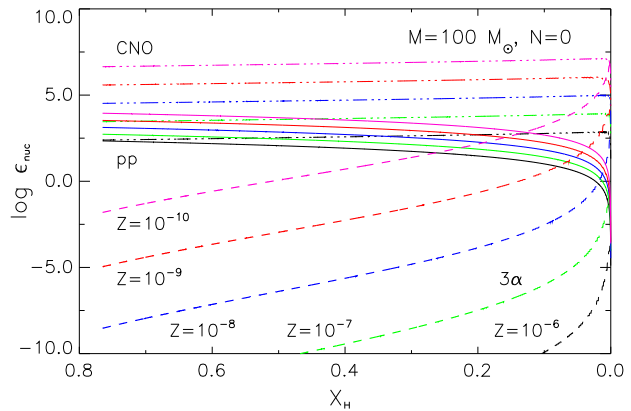}
\caption{Nuclear energy generation via pp-chain, CNO-cycle and
$3\alpha$-reaction for galactic and pregalactic Population III stars
with initial metallicity $Z=10^{-6}$ to $Z=10^{-10}$, without mass
loss during the hydrogen burning.}\label{bar_fig15}
\end{center}
\end{figure}

\begin{figure}
\begin{center}
\includegraphics [width=82mm]{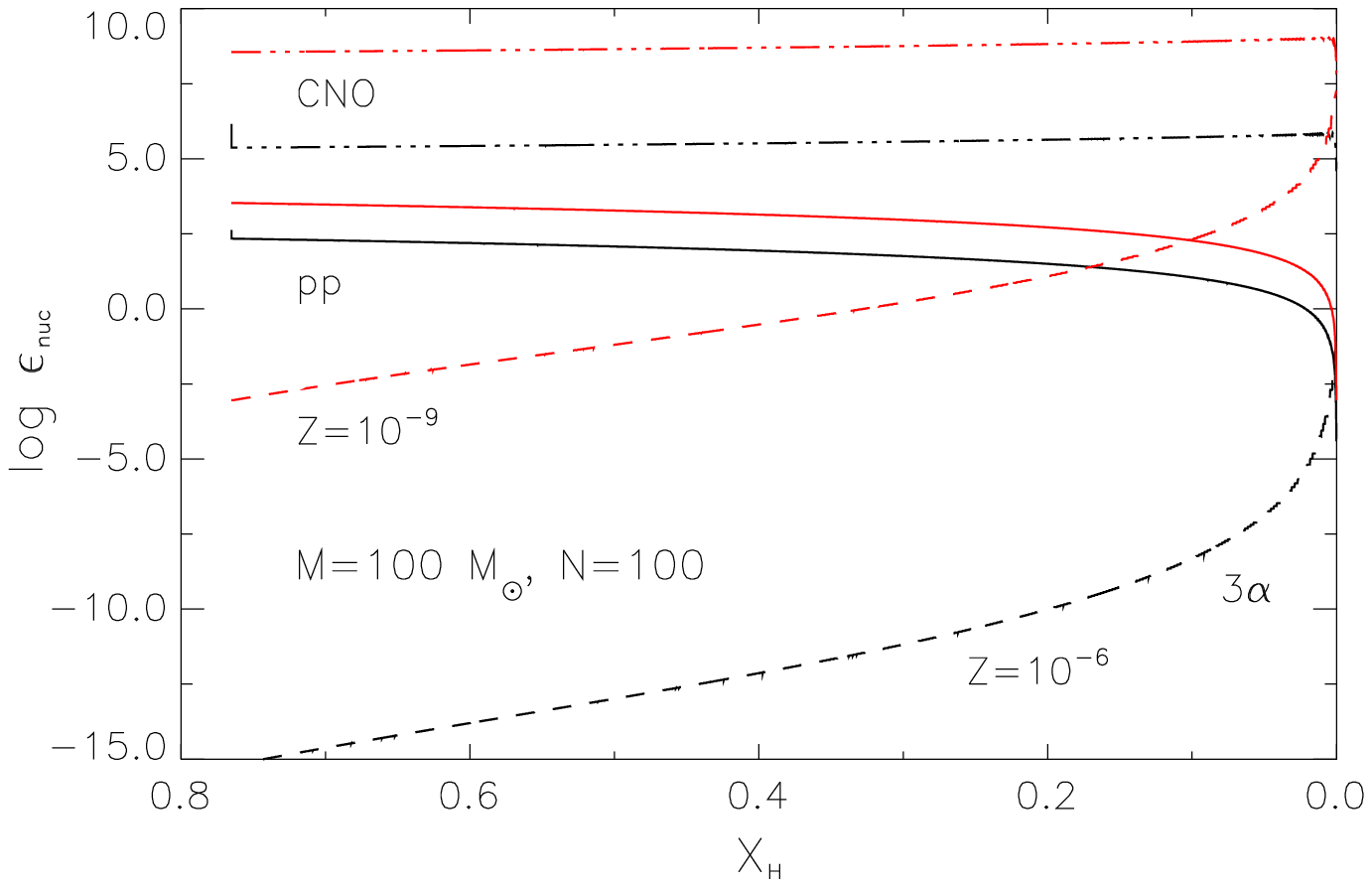}
\caption{Nuclear energy generation via pp-chain, CNO cycle and
$3\alpha$-reaction for galactic and pregalactic Population III stars
with initial metallicity $Z=10^{-6}$ and $Z=10^{-9}$, with mass loss
during the hydrogen burning.}\label{bar_fig16}
\end{center}
\end{figure}

 In our models, the lower is the metallicity, the sooner is activated
the triple-$\alpha$ reaction during the main-sequence.
Lower-metallicity pregalactic stars without mass loss produce higher
nuclear energy via the triple-$\alpha$ reaction from the beginning
of the main-sequence. For galactic stars contribution of this
reaction is at the end of the main-sequence only. This is shown in
Figure~\ref{bar_fig15}. For mass losing stars, the nuclear energy
production via the pp-chain, CNO-cycle and 3$\alpha$ reaction is
higher than in the conservative case, and 3$\alpha$ reaction starts
from the beginning of the main-sequence, as it is shown in
Figure~\ref{bar_fig16}.

 As a result of lower energy-generation rate in the core, the stars
maintain high central temperatures to support the stellar mass
against the gravitational collapse. This high temperatures in the
core and reduced opacity in their envelopes make these stars hotter
and more compact than their enriched counterparts, depending on the
initial stellar mass and metallicity.

 The lower is the metallicity, the more compact and hotter are the stars.
Table 1 shows models with $100\,M_{\odot}$ when they settle down on
the main-sequence. A galactic star with metallicity $Z=10^{-6}$ has
a radius of $7.68\,R_{\odot}$, while a pregalactic star with
$Z=10^{-9}$ has $4.49\,R_{\odot}$ or with $Z=10^{-10}$ it has
$3.69\,R_{\odot}$. Concerning the effective temperature this is
69582, 91865 and 101800 K, respectively. On the ZAMS, the
pregalactic lower metallicity stars are denser, hotter and slightly
more luminous than galactic ones.

 With effective temperatures of $T_{\rm eff}\sim 10^{5}$ K these
stars are very efficient at producing photons capable of ionizing
the hydrogen and helium \citep{Bromm2001}.

\begin{figure}
\begin{center}
\includegraphics [width=82mm]{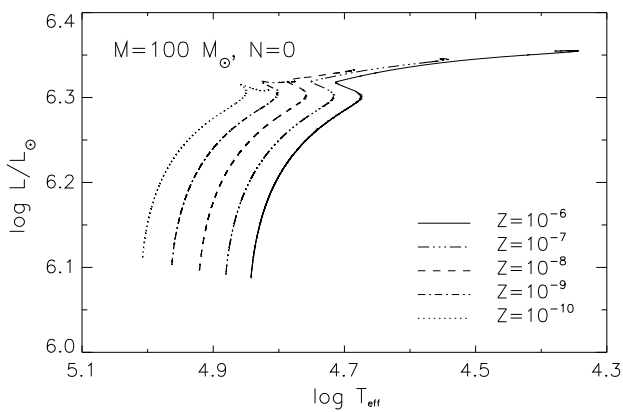}
\caption{Evolutionary tracks in the HR-diagram for a $100 M_{\odot}$
stars with different low-metallicity at constant mass. The lower is
the metallicity the bluer are the loci of the stars.}\label{bar_fig17}
\end{center}
\end{figure}

\begin{figure}
\begin{center}
\includegraphics [width=82mm]{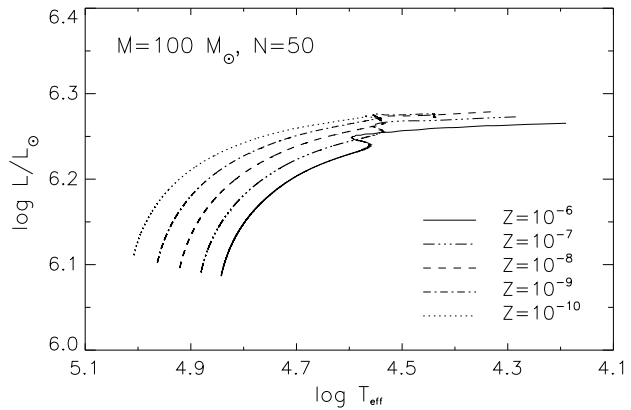}
\caption{Idem dito. Evolutionary tracks in the HR-diagram for a $100
M_{\odot}$ stars with different low-metallicity for mass losing
stars by using a mass loss parameter $N=50$.}\label{bar_fig18}
\end{center}
\end{figure}

 Lower metallicity stars are shifted to the blue upper part of the
HR-diagram. When stars are gradually enriched with heavy elements
they evolve redwards. This is shown in Figure~\ref{bar_fig17} for a
$100\,M_{\odot}$ model without a mass loss. Mass losing stars evolve
reducing their mass, luminosity and effective temperature, as it is
shown in Figure~\ref{bar_fig18}.

 In our models, evolutionary tracks describe different paths
on the HR-diagram depending on stellar mass and mass loss rates.
Galactic and pregalactic Population III mass losing stars move from
the left to the right in the upper part of this diagram. With higher
mass loss rates, the luminosity and the effective temperature
decreases more than in the conservative case.

\begin{table}
\small \caption{Some characteristics of lower-metallicity
$100\,M_{\odot}$ stars when settle down on the main-sequence. Models
become more hotter, denser, luminous and compact, as metallicity is
decreasing.\label{tbl-1}}
$$
\begin{tabular}{@{}cccccc@{}}
\tableline \small $Z$ & log $T_{\rm c}$ & log $\rho_{\rm c}$ & log
$\displaystyle{\frac{L}{L_{\odot}}}$ & log $T_{\rm
eff}$ & log $R$ \\
\hline
10$^{-6}$  & 7.84650 & 0.87274 & 6.08715 & 4.84250 & 11.72763 \\
10$^{-7}$  & 7.92150 & 1.09791 & 6.09132 & 4.88074 & 11.65323 \\
10$^{-8}$  & 8.00027 & 1.33451 & 6.09652 & 4.92088 & 11.57558 \\
10$^{-9}$  & 8.08323 & 1.58381 & 6.10314 & 4.96315 & 11.49433 \\
10$^{-10}$ & 8.17087 & 1.84730 & 6.11174 & 5.00775 & 11.40942 \\
\hline
\end{tabular}
$$
\end{table}

 As a result of their high mass and temperature, very massive stars
are dominated by radiation pressure and they have luminosities
closer to the Eddington limit. The specific luminosity $L_{\nu}\sim
L_{\rm Edd}\sim M$, and the total luminosity depends on the total
mass of stars only but it is independent of the IMF
\citep{Bromm2001}, which is a unique characteristic of very massive
stars.

 Mass loss can affect the stellar evolution in a variety of ways. In
the present work, nuclear lifetimes for mass losing stars with a
mass loss parameter $N=50$ are similar to those in the conservative
case.

 Hydrogen and helium burning lifetimes are decreasing with increasing
stellar mass. On the other hand, the nuclear lifetimes are
increasing with increasing mass loss rates. Nuclear lifetimes also
are different depending on the initial chemical composition of the
stars. For $M\geq 100\,M_{\odot}$ galactic and pregalactic
Population III stars the nuclear lifetimes during the hydrogen
burning are $\tau_{\rm H}\leq 3\times 10^{6}$ years. The helium
burning lifetimes are shorter than the hydrogen burning lifetimes.

 As discussed in Section~\ref{massloss}, it is still an open question
if the lower metallicity stars could have lost a substantial part of
their mass and which could be the mechanisms. We argue that a mass
loss is necessary to have a good mixing and transport of the
produced elements to the surface of the star. In the conservative
case, without any mass loss, the mixing is very deficient and the
CNO elements are not transported to the surface. However, in a
scenario with mass loss, the mixing is very efficient. The main
result is a production of N$^{14}$, which originates primarily in
the pregalactic stars, and subsequently in the galactic ones
\citep{Bahena2006}, as it will be shown in details in a forthcoming
paper.

 \cite{Meynet2002} and \cite{MaederMeynet2003} have shown
that the stellar rotation and mass loss can affect the predictions
of the stellar yields especially for He, C, N and O. At low $Z$, the
rotation may be a dominant effect in massive star evolution. They
estimate that low $Z$ stars may loose mass due to fast rotation as a
consequence of their initial slow winds, and initial rotation
velocities may be faster \citep{Maeder1999}. Their models of massive
stars at low $Z$ do not reach in general the red stage. However,
rotation permits the envelope to largely expand and the star moves
to the red supergiant stage, modifying all the models outputs.
 Then, the rotation affects the evolution of very metal-poor massive
stars \citep{Meynet2005} and can induce mass loss in many ways
\citep{Meynet2006a}. The compactness of low metallicity stars is a
key property which has many effects in relation with stellar
rotation.

\begin{table*}
\small \caption{Nuclear life times and masses during hydrogen and
helium burning for $100$, $120$, $150$, $200$ and $250\,M_{\odot}$
galactic Population III stars with initial metallicity $Z=10^{-6}$,
and diverse mass loss parameters.\label{tbl-2}}

$$
\begin{tabular}{@{}cccccccccccc@{}}
\hline
\\
$M_{\rm i}$ & $N$ & $\tau_{\rm H}$ & $\Delta M_{\rm H}$ & $q_{\rm
cc, H}$ & $M_{\rm f, H}$ & $M_{\rm He}$ & $\tau_{\rm He}$ & $q_{\rm
cc, He}$ & $M_{\rm f, He}$ & $M_{\rm CO}$ \\
\\
\hline
 100 & \ \ \, 0 & 3.09834 &\ \ 0.00000 & 0.41562 &  100.00000 &\ 41.56200 & 3.14857 & 0.38515 & 100.00000 &\ 38.51500 \\
     &     \ 50 & 3.07066 &   14.64268 & 0.45946 & \ 85.35732 &\ 39.21827 & 3.07857 & 0.33416 &\ 85.30596 &\ 28.50584 \\
     &      100 & 3.15136 &   26.82291 & 0.45071 & \ 73.17709 &\ 32.98165 & 3.15890 & 0.30761 &\ 73.09620 &\ 22.48512 \\
 120 & \ \ \, 0 & 2.89075 &\ \ 0.00000 & 0.43025 &  120.00000 &\ 51.64200 & 2.93901 & 0.40415 & 120.00000 &\ 48.49800 \\
     &     \ 50 & 2.87198 &   17.78249 & 0.47813 &  102.21751 &\ 48.87326 & 2.87981 & 0.33142 & 102.21992 &\ 33.87773 \\
     &      100 & 2.94026 &   32.84446 & 0.47004 & \ 87.15554 &\ 40.96659 & 2.94880 & 0.34399 &\ 87.04112 &\ 29.94127 \\
 150 & \ \ \, 0 & 2.67959 &\ \ 0.00000 & 0.44097 &  150.00000 &\ 66.14550 & 2.72591 & 0.42386 & 150.00000 &\ 63.57900 \\
     &     \ 50 & 2.67557 &   23.56230 & 0.49297 &  126.43770 &\ 62.32999 & 2.68253 & 0.38855 & 126.42807 &\ 49.12363 \\
     &      100 & 2.73644 &   43.49372 & 0.49057 &  106.50628 &\ 52.50440 & 2.74280 & 0.38651 & 106.33505 &\ 41.09956 \\
 200 & \ \ \, 0 & 2.46549 &\ \ 0.00000 & 0.45237 &  200.00000 &\ 90.47400 & 2.50979 & 0.42152 & 200.00000 &\ 84.30400 \\
     &     \ 50 & 2.43380 &   32.71329 & 0.50490 &  167.28671 &\ 84.46306 & 2.48056 & 0.37267 & 167.27864 &\ 62.33973 \\
     &      100 & 2.51988 &   59.07631 & 0.50571 &  140.92369 &\ 71.26652 & 2.52582 & 0.34120 & 140.68414 &\ 48.00143 \\
 250 & \ \ \, 0 & 2.32990 &\ \ 0.00000 & 0.46402 &  250.00000 & 116.00500 & 2.37348 & 0.45782 & 250.00000 & 114.45500 \\
     &     \ 50 & 2.34590 &   41.44418 & 0.51272 &  208.55582 & 106.00500 & 2.35321 & 0.40578 & 208.37607 &\ 84.55484 \\
     &      100 & 2.38727 &   45.41756 & 0.51194 &  174.58244 &\ 89.37573 & 2.39382 & 0.41550 & 174.27963 &\ 72.41319 \\
\hline
\end{tabular}
$$
\end{table*}

 For the present work, Tables 2 and 3 indicate some features for
galactic and pregalactic Population III mass losing stars. The first
column lists the initial stellar mass $M_{i}$, the second one refers
to the mass loss parameter $N$. The next columns indicate the
lifetime $\tau_{\rm H}$ during hydrogen burning given in Mega-years,
the mass ejecta during the hydrogen burning $\Delta M_{\rm H}$, the
convective core size $q_{\rm cc,\,H}$ during this phase and the
corresponding He-core mass $M_{\rm He}$. Then there follow the
lifetime $\tau_{\rm He}$ during helium burning in Mega-years, the
convective core size $q_{\rm cc,\,He}$ during this burning phase and
the CO-core mass $M_{\rm CO}$. The mass ejecta, the He- and CO-core
masses are given in solar masses.

 In our models the mass ejecta are important during hydrogen burning
and they increase with increasing initial stellar mass. For $100$ and
$200\,M_{\odot}$ pregalactic stars, with $N=100$, the ejecta $\Delta
M_{\rm H}$ are $26.82$ and $59.08\,M_{\odot}$, respectively.

 In the present work, mass losing stars form a convective core size
slightly greater than in the conservative case. Stars evolving with
mass loss  parameter $N=100$ form a helium core $M_{\rm He}=32.98$,
and $71.27\,M_{\odot}$ for $100$ and $200\,M_{\odot}$ galactic
($Z=10^{-6}$) stars, respectively. For pregalactic ($Z=10^{-9}$)
stars, $M_{\rm He}=32.29$ and $68.72\,M_{\odot}$, respectively. The
corresponding carbon core masses are $22.48$ and $48.00\,M_{\odot}$
for the first case, and $32.13$ and $70.24\,M_{\odot}$ for the
second one, respectively.

 Studies of core-collapse SNe have discovered two
distinct types of supernovae explosions: (i) very energetic SNe or
hypernovae (HNe), and (ii) very faint and low energy SNe. In
\cite{Nomoto2004} it was proposed that the first generation
supernovae were the explosions of $\sim 20-30\,M_{\odot}$ stars and
some of them produced C-rich, Fe poor ejecta. The ejecta of these
explosions can well account for the abundance pattern of EMP stars.
In contrast, the observed abundance patterns cannot be explained by
the explosions of more massive, $130-300\,M_{\odot}$ stars. These
stars undergo pair-instability SNe explosions and are disrupted
completely \citep{Umeda2002, Heger2002}.

\begin{table*}
\small \caption{Nuclear life times and masses during hydrogen and
helium burning for $100$, $120$, $150$, $200$ and $250\,M_{\odot}$
pregalactic Population III stars with initial metallicity
$Z=10^{-9}$, and diverse mass loss parameters.\label{tbl-3}}

$$
\begin{tabular}{ccccccccccccccc}
 \hline
 \\
 $M_{\rm i}$ & $N$ & $\tau_{\rm H}$ & $\Delta M_{\rm H}$ & $q_{\rm cc, H}$
 & $M_{\rm f, H}$ & $M_{\rm He}$ & $\tau_{\rm He}$ & $q_{\rm cc, He}$
 & $M_{\rm f, He}$ & $M_{\rm CO}$ \\
 \\
 \hline
 100 & \ \ \, 0 & 2.89133 &\ \ 0.00000 & 0.37951 & 100.00000 &\ 37.95100 & 2.93547 & 0.38910 & 100.00000 &\ 38.91000 \\
     &     \ 50 & 2.89685 &   14.31142 & 0.44794 &\ 85.68958 &\ 38.38334 & 2.93011 & 0.44891 &\ 85.39561 &\ 38.33494 \\
     &      100 & 2.95916 &   26.23277 & 0.43772 &\ 73.76723 &\ 32.28939 & 2.97677 & 0.43679 &\ 73.55861 &\ 32.12967 \\
 120 & \ \ \, 0 & 2.69832 &\ \ 0.00000 & 0.38496 & 120.00000 &\ 46.19520 & 2.73923 & 0.40604 & 120.00000 &\ 48.72480 \\
     &     \ 50 & 2.70386 &   17.72365 & 0.46846 & 102.27635 &\ 47.91238 & 2.74578 & 0.46763 & 102.26871 &\ 47.82392 \\
     &      100 & 2.76936 &   32.99806 & 0.46448 &\ 87.00194 &\ 40.41066 & 2.81274 & 0.46295 &\ 86.43422 &\ 40.01472 \\
 150 & \ \ \, 0 & 2.50358 &\ \ 0.00000 & 0.39613 & 150.00000 &\ 59.41950 & 2.54207 & 0.41884 & 150.00000 &\ 62.82600 \\
     &     \ 50 & 2.52162 &   22.89308 & 0.48177 & 127.10692 &\ 61.23630 & 2.52587 & 0.49071 & 127.09890 &\ 62.36826 \\
     &      100 & 2.57387 &   41.94500 & 0.47805 & 108.05500 &\ 51.65569 & 2.57775 & 0.49219 & 108.04709 &\ 53.17970 \\
 200 & \ \ \, 0 & 2.30257 &\ \ 0.00000 & 0.41782 & 200.00000 &\ 83.56400 & 2.34043 & 0.43612 & 200.00000 &\ 87.22400 \\
     &     \ 50 & 2.34086 &   31.96341 & 0.48552 & 168.03659 &\ 81.58513 & 2.34750 & 0.48872 & 168.02637 &\ 82.11785 \\
     &      100 & 2.38205 &   58.43802 & 0.48546 & 141.56198 &\ 68.72268 & 2.38542 & 0.49668 & 141.41945 &\ 70.24021 \\
 250 & \ \ \, 0 & 2.17765 &\ \ 0.00000 & 0.42354 & 250.00000 & 105.88500 & 2.21375 & 0.44310 & 250.00000 & 110.77500 \\
     &     \ 50 & 2.22612 &   40.76518 & 0.48886 & 209.23482 & 102.28653 & 2.23255 & 0.49352 & 209.16652 & 103.22786 \\
     &      100 & 2.25313 &   74.29857 & 0.48770 & 175.70143 &\ 85.68959 & 2.25955 & 0.49271 & 175.39363 &\ 86.41820 \\
\hline
\end{tabular}
$$
\end{table*}

 If all stars evolve with mass loss they can reduce their initial
mass to massive stars range, as our models show. Depending of their
initial mass and mass loss rate, its final fate could be either SNe
or HNe. The HNe explosions connected to some low-$z$ gamma-ray
bursts could be relevant to early nucleosynthesis at low metallicity
\citep{Umeda2002}.


\section{Conclusions}\label{conclusions}


 The main aim of the present study was to construct models for very
massive low-metallicity both galactic and pregalactic Population III
mass losing stars. We have calculated and discussed here the main
properties of these stars in comparison with the conservative case
presented in Paper I. Our results show that the mass loss could be
very important for massive low metallicity stars, although the
physics and mechanisms of the mass loss are not well known yet.
According to the line-driven wind theory, the mass loss rate is
negligible for the very low metallicity. However, the mass loss may
occur due to other processes which are metallicity independent, such
as effects of continuum radiation, pulsation and/or rotation etc. In
our approach we have chosen an ad hoc rate of the mass loss and we
used the know physics to explore its effects on the stellar
evolution at very low $Z$.

 For evolution with mass loss, the reduction of the stellar mass
makes the central temperature to increase less rapidly than for
constant-mass evolution. Thus the mass of the convective core
decreases more rapidly as the evolution proceeds. However, the core
mass fraction is larger in a star evolving with mass loss. The
luminosity of a star evolving with mass loss is lower than for
constant-mass evolution and in consequence its main-sequence
lifetime is somewhat increased.

 The evolution of VMS both with and without mass loss takes place in
the upper part of the HR-diagram. In the absence of heavy CNO elements,
lower metallicity galactic and pregalactic stars contract and settle down
on the ZAMS with central temperatures high enough for some $^{4}$He to be
converted to CNO nuclides through the triple-$\alpha$ reaction and
CNO cycles. The energy generated stops the contraction and the star
re-expands.

 In all cases, lower metallicity pushes the stars to a smaller radius
at higher density and temperature. Therefore, galactic and
pregalactic Population III stars are more compact, hotter and denser
than their metal enriched counterparts. Due to their different
chemical composition, the luminosity is slightly lower for
pregalactic stars.

 The evolution of both galactic and pregalactic Population III stars
is influenced by the increase of the mean molecular weight in the
convective core and the effect of mass loss. The mean molecular
weight affects the star by increasing their luminosity. As a result
of the increased mean molecular weight in the convective core the
central temperature also increases. With increased molecular weight
the convective core shrinks. The contraction of the convective core
makes the star to expand and the effective temperature to decrease.
The proportionality between mass and luminosity causes that the
luminosity decreases as the star loses mass.

 \cite{Heger2002} and \cite{Hegeretal2002, Heger2003} have studied
the final fate of VMS considering only the conservative evolution.
Stars with $140\,M_{\odot}<M<260\,M_{\odot}$ explode as
Pair-Instability Supernovae Explosions (PISNe) causing complete
disruption. In these cases, $M_{\rm He}\sim 64-133\,M_{\odot}$.

 Our models for $200\,M_{\odot}$ galactic and pregalactic stars
develop in the conservative case $M_{\rm He}\sim 90$ and
$84\,M_{\odot}$, respectively. With moderate mass loss rate
($N=50$), they develop $M_{\rm He}\sim 84$, and $82\,M_{\odot}$.
With increasing mass loss rate ($N=100$), stars develop $M_{\rm
He}=71$ and $69\,M_{\odot}$. Then, these galactic and pregalactic
stars could explode as PISNe.

 However, according to scenarios by \cite{Umeda2002, Umeda2003,
Umedaetal2002, Nomoto2004}, the stars with $M<130\,M_{\odot}$ could
explode like HNe. Our $100\,M_{\odot}$ galactic and pregalactic
models develop $M_{\rm He}\sim 42$ and $38\,M_{\odot}$ in the
conservative case ($N=0$), $M_{\rm He}\sim 39$ and $38\,M_{\odot}$
with a moderate mass loss rate ($N=50$), and with $N=100$, they have
$M_{\rm He}=33$ and $32\,M_{\odot}$, respectively. Then, models for
$\leq 100\,M_{\odot}$ mass losing stars, which evolve with high mass
loss rates, can reduce appreciably their initial mass during the
hydrogen, and helium burning. Their cores could also explode like
HNe, which are likely to make an important contribution to the early
galactic chemical evolution.

 The most important effect in the transition in the star formation
properties from the first stars to present day is the presence of a
trace amount of heavy elements. The concept of critical metallicity
has been used to characterize the transition between Population III
and Population II star formation modes \citep{Omukai2001,
Bromm2001,Bromm2004}. This important parameter value of critical
threshold for enabling a formation of stars with lower mass is
estimated to be $Z\sim 10^{-3.5}Z_{\odot}\sim 6.32\times 10^{-6}$.
In our very massive evolutionary models, their initial mass is
reduced when they lost mass reaching the range of massive stars. Due
to the mass loss an efficient mixing of CNO elements is reached.
After hydrogen burning, the initial metallicity increases providing
an initial enrichment of the IGM with heavy elements, giving place
to the next generation of stars with increasing metallicity. In
successive generations the EMP stars observed in present days could
be formed.

 The evolution of the first stars with mass loss is still an
open problem. It would be interesting to suppose that the rotation
or other mechanisms driving the mass loss is simultaneously
responsible for a mixing of elements in the stellar interiors. The
main problem, however, is to find a realistic model for the mass
loss rate due to the different mechanisms and their combinations and
to include it into self-consistent models of the stellar evolution.


\begin{acknowledgements}
\small This work has been partially supported by the Mexican Consejo
Nacional de Ciencia y Tecnolog\'{\i}a (CONACyT). DB thanks to
CONACyT for a Ph. D. grant, P.H. acknowledges grants LC0614 and
202/09/0772.
\end{acknowledgements}


\bibliographystyle{spr-mp-nameyear-cnd}
\bibliography{biblio-u1}

\label{lastpage}

\end{document}